\documentclass[aps,final,notitlepage,oneside,twocolumn,nobibnotes,nofootinbib,superscriptaddress,
noshowpacs,centertags]{revtex4-1}

\usepackage[english]{babel}
\usepackage{graphicx}
\usepackage{latexsym}
\usepackage{amssymb}
\usepackage{amsmath}

\begin{document}

\title{Stationary Solutions of the Dirac Equation in the Gravitational Field
of a Charged Black Hole}
\author{V.I. Dokuchaev}\thanks{e-mail: dokuchaev@inr.ac.ru}
\author{Yu.N. Eroshenko}\thanks{e-mail: eroshenko@inr.ac.ru}
\affiliation{Institute for Nuclear Research, Russian Academy of Sciences, prosp. 60-letiya Oktyabrya 7a, Moscow, 117312 Russia}

\date{\today}

\begin{abstract}
A stationary solution of the Dirac equation in the metric of a Reissner-Nordstr\"om black hole has
been found. Only one stationary regular state outside the black hole event horizon and only one stationary
regular state below the Cauchy horizon are shown to exist. The normalization integral of the wave functions
diverges on both horizons if the black hole is non-extremal. This means that the solution found can be only
the asymptotic limit of a nonstationary solution. In contrast, in the case of an extremal black hole, the normalization integral is finite and the stationary regular solution is physically self-consistent. The existence of
quantum levels below the Cauchy horizon can affect the final stage of Hawking black hole evaporation and
opens up the fundamental possibility of investigating the internal structure of black holes using quantum tunneling between external and internal states.
\end{abstract}

\maketitle


\bigskip

\section{Introduction}

The Dirac equation in a gravitational field of general form was derived by Fock and Iwanenko in 1929
\cite{Fock29} using the formalism of parallel spinor transport,
which allowed its covariant derivative to be determined. Another method for deriving the Dirac equation in a gravitational field is based on the description
of the Lorentz group in tetrad formalism \cite{Wei00}. The
Dirac equation written in a (pseudo-)Riemannian
space is
\begin{equation}
(i\gamma^\mu D_\mu-m)\psi=0,
\label{urd}
\end{equation}
where the Dirac matrices in a metric of general form
$\gamma^\mu=e^{\phantom{00}\mu}_{(a)}\gamma^{(a)}$ are expressed in terms of the standard
matrices $\gamma^{(a)}$ in a Minkowski space via the tetrad $e^{\phantom{00}\mu}_{(a)}$.
The extended derivative is
\begin{equation}
D_\mu=\partial_\mu+iqA_\mu+\Gamma_\mu,
\end{equation}
where
\begin{equation}
\Gamma_\mu=\frac{1}{4}\gamma^{(a)}\gamma^{(b)}e^{\phantom{00}\nu}_{(a)}e_{(b)\nu\,;\mu},
\end{equation}
and $A_\mu$ is the electromagnetic 4-potential.

The Dirac equation in gravitational fields of various forms was investigated in many papers (see, e.g.,
\cite{BriWhe57,Muk00,Ber75}). The scattering of fermions by black holes and
the emission of fermions in the process of quantum
black hole evaporation (the Hawking effect) were
studied most extensively. Most of the efforts at investigating the Dirac equation in a gravitational field were
focused on problems of this type. The resonant quasistationary quantum states of scalar particles in the
gravitational field of a black hole were investigated in
\cite{DerRuf74,Teretal78,Kof82}. Similar quantum states for spinor particles were
analyzed in \cite{SofMulGre77,Teretal80,GalPomChi83}. The stationary states of charged
particles were studied in the gravitational field of a
Schwarzschild black hole \cite{GorNez12}, \cite{Vroetal13} and an electrically
charged Reissner-Nordstr\"om black hole \cite{Dzh12} in a
region outside the horizon.

The spacetime of an eternal Reissner-Nordstr\"om
black hole is an infinite sequence of internal universes
\cite{chandra}. A particle falling into an eternal Reissner-Nordstr\"om black hole can either escape into another internal universe or be left below the Cauchy horizon. The
existence of stable finite orbits for classical particles
below the Cauchy horizon was shown in \cite{Dok11,BalBicStul89,Kagramanova10,GrunKag10,Olietal11}.

Here, we consider the stationary quantum states of
fermions inside and outside a black hole and show that
such states and the corresponding energy levels actually exist under certain conditions. A Reissner-Nordstr\"om black hole with a charged particle at a quantum
level in some respect resembles the simplest hydrogen
atom. However, in the case of a charged black hole, the
electron levels can be not only outside but also inside
the black hole, below the Cauchy horizon. In addition, the boundary conditions for the wave functions
on the black hole horizon are distinctly different,
which changes qualitatively the energy level characteristics for stationary states. The existence of energy levels inside a black hole opens up the fundamental new
possibility of ``looking'' into the black hole using
quantum methods, which is impossible within the
framework of classical general relativity. More specifically, if there exist energy levels inside a black hole,
then the internal structure of black holes can be studied based on the spectrum of transitions between
external and internal levels.

The ultimate fate of black holes evaporating in the
Hawking process has not yet been clarified, in particular, because large deviations from predictions of the
classical gravitation theory are probable as the black
hole mass approaches the Planck mass $M_{\rm Pl}=\sqrt{\hbar c/G}\approx10^{-5}$~g. The various effects that could stabilize an evaporating primordial black hole near the mass$M_{\rm Pl}$ were
discussed (for an overview of the models, see \cite{CarGilLid94}).
These remnants of evaporating black holes, called
``planckions'', were proposed as candidate dark matter
(hidden mass) particles in the Universe \cite{DolNasNov00}, \cite{Carr03}. Black
holes of such masses should have distinct quantum
properties and, therefore, the quantum states of particles we discuss can play a prominent role in the properties of these black holes, in particular, they can
change the probability of their quantum decay or final
evaporation. Previously, Markov discussed the so-called maximons, semi-closed worlds that are particle-like charged solutions in general relativity \cite{Mar66}.
Here, we show that a new type of systems remotely
resembling maximons, charged black holes with
charges on inner quantum orbits, can exist. These systems can also represent dark matter if they are stable
and were born in sufficient quantities at early cosmological epochs.

Although eternal black holes with internal spaces
are probably not formed through classical gravitational collapse, they can emerge in quantum processes
during particle collisions on accelerators (experiments
on the Large Hadron Collider have already tested
some of such models) or when ultrahigh-energy cosmic ray particles interact with the atmosphere (these
interactions are observed experimentally on several
detectors) if the theories with extra space dimensions
are realized \cite{Carr03}. Since such mini-black holes are
formed in processes involving charged particles, it is
natural to expect the birth of not neutral but charged
Reissner-Nordstr\"om black holes with various configurations of charges on orbits inside and outside
these black holes. In this case, not the gradual collapse
of matter and the capture of particles into a black hole
but the instantaneous quantum birth of systems with
an internal structure will occur.

\section{The Dirac equation in the Reissner-Nordstr\"om metric}

Let us briefly describe the method of separation of
variables in the Dirac equation \cite{BriWhe57} as applied to a
Reissner-Nordstr\"om black hole with the metric
\begin{equation}
 \label{RN0}
 ds^2=fdt^2-f^{-1}dr^2-r^2(d\theta^2+\sin^2\!\theta\,d\phi^2),
\end{equation}
where $M$ is the mass of the black hole, $Q$ is its charge,
and $f=1-2M/r+Q^2/r^2$. We use the units of measurement in which $c=G=\hbar=1$. In the case of $|Q|<M$,
the equation $f(r)=0$ has two roots $r=r_\pm=M\pm\sqrt{M^2-Q^2}$,
the event horizon and the Cauchy horizon (internal
horizon). In the case of $|Q|>M$, metric (\ref{RN0}) describes
a naked singularity without an event horizon, while
$|Q|=M$ corresponds to an extremal black hole. The
electromagnetic potential for a static black hole with
charge $Q$ is $A_\mu=(Q/r,0,0,0)$. Using the tetrad
\begin{equation}
e^{(a)}_{\phantom{00}\mu}={\rm diag}(f^{1/2},f^{-1/2},r,r\sin\theta)
\end{equation}
the Dirac equation (\ref{urd}) takes the form
\begin{eqnarray}
&&\left[i\gamma^{(0)}\frac{1}{f^{1/2}}\frac{\partial}{\partial t}+i\gamma^{(1)}f^{1/2}\frac{\partial}{\partial r}+
i\gamma^{(2)}\frac{1}{r}\frac{\partial}{\partial \theta} \right.+
\nonumber \\
&+&\left. i\gamma^{(3)}\frac{1}{r\sin\theta}\frac{\partial}{\partial \phi}\right.  
+i\gamma^{(1)}\frac{Q^2-3Mr+2r^2}{2r^3f^{1/2}}+
\nonumber \\
&+&\left. i\gamma^{(2)}\frac{\cos\theta}{2r\sin\theta}-\gamma^{(0)}\frac{qQ}{rf^{1/2}}-m
\right]\psi=0.
\end{eqnarray}

Following \cite{BriWhe57}, to simplify this equation further, we
will redefine the wave function
\begin{equation}
\psi=\frac{e^{-iEt}\Psi}{f^{1/4}r\sin^{1/2}\theta}
\label{psipsi}
\end{equation}
and single out the operator
\begin{equation}
K=\gamma^{(0)}\gamma^{(1)}\left(\gamma^{(2)}\frac{\partial}{\partial \theta}+\gamma^{(3)}\frac{1}{\sin\theta}\frac{\partial}{\partial \phi}\right),
\end{equation}
with integer eigenvalues $K\Psi=k\Psi$, where $k=0,~\pm1,~\pm2,~\dots$. Below, we will not be interested in the angular
part $Z(\theta,\phi)$ expressed in terms of spherical harmonics,
but we assume the integral of $|Z(\theta,\phi)|^2$ over the solid
angle to be equal to one. After these substitutions, the
system breaks up into two pairs of equivalent equations. Writing $\Psi$ as
\begin{equation}
\Psi=Z(\theta,\phi)\left[
\begin{array}{c}
g(r)I_2\\
ih(r)I_2
\end{array}
\right],
\label{gih}
\end{equation}
where $I_2$ is the column $(1,1)^T$, we obtain a system of
equations for the radial wave functions:
\begin{equation}
\frac{dg}{dr}-\frac{gk}{rf^{1/2}}+\frac{h}{f^{1/2}}\left[\frac{1}{f^{1/2}}\left(E-\frac{qQ}{r}\right)+m\right]=0,
\label{itogeq1}
\end{equation}
\begin{equation}
\frac{dh}{dr}+\frac{hk}{rf^{1/2}}-\frac{g}{f^{1/2}}\left[\frac{1}{f^{1/2}}\left(E-\frac{qQ}{r}\right)-m\right]=0.
\label{itogeq2}
\end{equation}
These equations can be easily derived from Eqs. (39) in
\cite{BriWhe57} if we set $e^\nu=e^{-\lambda}=f$ and transform the expression
for the electromagnetic potential. Because of the
divergence$f(r)\to \infty$ on the black hole horizon,
Eqs. (\ref{itogeq1}) and (\ref{itogeq2}) belong to the class of differential
equations with singular points.

The zeroth component of the fermionic field probability flux $j^\mu=\bar\psi\gamma^\mu\psi$, the probability density, can be
normalized to unity:
\begin{equation}
2\int\limits_0^{r_-}\frac{|g|^2+|h|^2}{f(r)}dr=1,
\label{normint}
\end{equation}
where we used Eqs.~(\ref{psipsi}) and (\ref{gih}) and performed the integration $\int j^0\sqrt{\gamma}d^3r$ with the spatial metric tensor $\gamma_{\alpha\beta}$. The
probability flux along the radius $j^1=2i(g^*h-gh^*)$
becomes zero, in particular, for real $g$ and $h$ when there
are no traveling waves in the solution.


\section{A non-extremal black hole with $|Q|<M$}

We have $f\to Q^2/r^2$ at $Q\neq0$ near the singularity
$r\to0$ and the system of equations (\ref{itogeq1}), (\ref{itogeq2}) for $k=0$ has an asymptotic solution,
\begin{equation}
g=C_1+C_2\frac{q}{Q}\frac{r^2}{2}, \quad h=-C_1\frac{q}{Q}\frac{r^2}{2}+C_2,
\label{asc1c2}
\end{equation}
where $C_1$ and $C_2$ are constants. If $k\neq0$, then (\ref{itogeq1}) and
(\ref{itogeq2})  take the form $dg/dr=kg/Q$ and $dh/dr=-kh/Q$
and their asymptotic solutions are
\begin{equation}
g\propto\exp(kr/Q),~~~h\propto\exp(-kr/Q).
\label{asexp}
\end{equation}
Thus, the solutions enter into the singularity with zero
and finite derivatives for $k=0$ and $k\neq0$, respectively.
Solutions (\ref{asc1c2}) and (\ref{asexp}) are valid both for non-extremal black holes and for extremal ones and naked singularities.

Let us first consider the region below the internal
Cauchy horizon $r<r_-$. According to the classification
by I.D.~Novikov, this is the $R$-region. The local structure of the spacetime in it is the same as that outside
the black hole; in particular, $t$ and $r$ have the meaning
of time and radial coordinates. As was shown in \cite{Dok11,BalBicStul89,Kagramanova10,GrunKag10,Olietal11}, classical orbits of charged particles can exist at
$r<r_-$. In this section, we study the question about
the quantum orbitals of particles below the Cauchy
horizon.

To investigate the wave functions when $r\to r_-$, it is
convenient to introduce a new variable $y=(1-r/r_-)^{1/2}$.
Equations (\ref{itogeq1}) and (\ref{itogeq2}), to within $O(y^2)$, will then take the form
\begin{equation}
y\frac{dg}{dy}+2kpyg-2r_-ph\left[\left(E-\frac{qQ}{r_-}\right)p+my\right]=0,
\label{itogeq1y}
\end{equation}
\begin{equation}
y\frac{dh}{dy}-2kpyh+2r_-pg\left[\left(E-\frac{qQ}{r_-}\right)p-my\right]=0,
\label{itogeq2y}
\end{equation}
where $p=(r_+/r_--1)^{-1/2}$. Let us prove that a regular
solution of system (\ref{itogeq1y}), (\ref{itogeq2y}) exists for $r\to r_-$ only in the case
\begin{equation}
E=\frac{qQ}{r_-}.
\label{energy}
\end{equation}
We write $g$ and $h$ in the form of series,
\begin{equation}
g=y^s\sum\limits_{n=0}^{\infty}a_ny^n,~~~h=y^w\sum\limits_{n=0}^{\infty}b_ny^n,
\label{sergh}
\end{equation}
where $a_0\neq0$ and $b_0\neq0$. We reason from the contrary.
Suppose that $E\neq qQ/r_-$. If $s\neq0$ and $w\neq0$, then substituting (\ref{sergh}) into (\ref{itogeq1y}) and (\ref{itogeq2y}) and writing out the coefficients at the smallest powers of $y$, we obtain $s=w$ and
then
\begin{equation}
a_0s=2r_-p^2b_0\left(E-\frac{qQ}{r_-}\right),~~~b_0s=-2r_-p^2a_0\left(E-\frac{qQ}{r_-}\right).
\end{equation}
Multiplying these relations, we have
\begin{equation}
s^2=-4r_-^2p^4\left(E-\frac{qQ}{r_-}\right)^2.
\label{condx}
\end{equation}
Condition (\ref{condx}) can hold only if (\ref{energy}) is valid and $s=0$,
which contradicts our assumptions. Let now one of
the quantities $s$ and $w$ be equal to zero, for example, let
us set $s=0$. We then find from (\ref{itogeq1y}) that $w\ge1$. However, this value will contradict Eq.~(\ref{itogeq2y}), because the
power of $y$ at the coefficient $(E-qQ/r_-)$ is $s=0$, while
in the first term the exponent $w\ge1$. The case of $w=0$
is considered similarly. We again came to a contradiction. Consequently, the only condition under which
Eqs.~(\ref{itogeq1y}) and (\ref{itogeq2y}) are satisfied is specified by Eq.~(\ref{energy}).
Thus, the charge below the internal horizon of a Reissner-Nordstr\"om black hole in a stationary state $\propto e^{-iEt}$
can have only one fixed energy. Despite the fact that
the solution of system (\ref{itogeq1y}), (\ref{itogeq2y}) for $r\to r_-$ under
condition (\ref{energy}) formally exists, it cannot correspond to
a real physical situation due to the divergence of the
normalization integral (\ref{normint}) on the horizon $r_-$. Indeed,
exact solutions of Eqs.~(\ref{itogeq1y}) and (\ref{itogeq2y}) can be easily
obtained under condition (\ref{energy}):
\begin{equation}
g=C_1e^{\lambda y}+C_2e^{-\lambda y},
\end{equation}
where $\lambda=2p\sqrt{k^2+r_-^2m^2}$ and $h(r)$ is expressed in terms
of $g(r)$ from (\ref{itogeq1y}). For the convergence of (\ref{normint}), we
should choose $C_1=-C_2$, but then $h\to const$ when
$r\to r_-$. Similarly, we obtain $g\to const$ const when $h\to0$.
In both cases, (\ref{normint}) diverges.

Just as in the case of a Schwarzschild black hole
\cite{GorNez12}, system (\ref{itogeq1y}), (\ref{itogeq2y}) for $E\neq qQ/r_-$ and $y\to0$ has
an irregular solution,
\begin{equation}
g=C\sin(\alpha\ln y+\delta), \qquad h=C\cos(\alpha\ln y+\delta),
\label{nonreg1}
\end{equation}
where $C$ and $\delta$ are constants, $\alpha=2r_-p^2(E-qQ/r_-)$.
However, $|g|^2+|h|^2=|C|^2$ for this solution and (\ref{normint})
diverges on the horizon.

The solution at infinity $r\to\infty$ is the same as that
in the case of an ordinary hydrogen atom:
\begin{equation}
g=C_1e^{-ir\sqrt{E^2-m^2}}+C_2e^{ir\sqrt{E^2-m^2}},
\end{equation}
where $h=-(E+m)^{-1}dg/dr$. A localized, exponentially decreasing solution exists only for $|E|<m$, i.e.,
the particle–central charge interaction should have
the pattern of attraction with a negative contribution
to $E$.

To investigate the solution near the horizon $r\to r_+$,
let us designate $y=(r/r_+-1)^{1/2}$ and $p=(1-r_-/r_+)^{-1/2}$.
We analyze the corresponding equations just as (\ref{itogeq1y})
and (\ref{itogeq2y}) and find that the charge has only one energy
level,
\begin{equation}
E=\frac{qQ}{r_+},
\label{energyout}
\end{equation}
which was specified in \cite{Dzh12}. However, (\ref{normint}) again
diverges. The conclusion about the divergence of the
normalization integral of the external solution on the
horizon was also reached in \cite{Dzh12}.

Divergence on the horizons points to the possibility
of nonstationary solutions that are localized in the
course of time on the horizons, while their energy
tends to (\ref{energyout}) and (\ref{energy}). The resonant quasi-stationary
levels outside a black hole were investigated in a number of papers, where the difficulty with the behavior on
the horizon was circumvented by the passage to a new
variable $dr^*/dr=f^{-1}(r)$, which moves the horizon to
$r^*=-\infty$ (see \cite{SofMulGre77}). However, the strictly stationary
levels that we discuss here do not exist because of the
above divergence on the horizon.

The difference between atoms with a set of levels
and hydrogen-like atoms is attributable to the change
of the boundary conditions for the Dirac equation.
Quantization in the hydrogen atom follows from the
finiteness of the wave function or the normalization
probability integral. The presence of an event horizon
changes the form of the boundary conditions and
equations. As a consequence, only one energy level
exists for a strictly stationary solution.

\section{The solution for an extremal black hole}

\begin{figure}[t]
\begin{center}
\includegraphics[angle=0,width=0.49\textwidth]{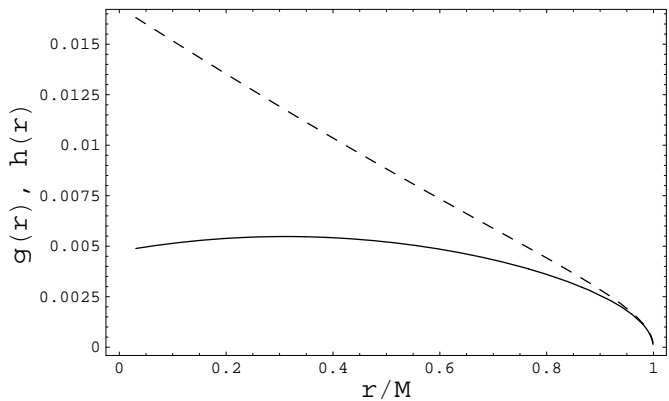}
\end{center}
\caption{Wave functions $g(r)$ (solid curve) and $h(r)$ (dashed curve)
below the horizon of an extremal black hole at $k=1$. As an
example, we chose the parameters $\mu=1$ and $\nu=-0.5$ and
normalized the solutions using (\ref{normint}).}
\label{gr1extr}
\end{figure}

In the case of an extremal black hole, $M=|Q|$,
both horizons coincide: $r_-=r_+=r_h=M$. This changes
the boundary condition and, as we will now show, creates conditions for the existence of a physically
acceptable solution. The solution at the center when
$r\to 0$ has the same asymptotics (\ref{asc1c2}) and (\ref{asexp}). When
$r\to r_-$ from the inside, we introduce a new variable
$y=f^{1/2}=(r_h/r-1)$. Equations (\ref{itogeq1}) and (\ref{itogeq2}) will then take the form
\begin{equation}                                     
\label{itogeq2exin}
 y^2(1+y)^2\frac{dh}{dy}-hky(1+y)+gr_h\!\left[E\!
 -\!\frac{qQ(1+y)}{r_h}\!-\!my\right]\!=\!0. 
\end{equation}
\begin{equation}
 y^2(1+y)^2\frac{dg}{dy}+gky(1+y)-hr_h\!\left[E\!
 -\!\frac{qQ(1+y)}{r_h}\!+\!my\right]\!=\!0,
\label{itogeq1exin}
\end{equation}
The combinations of quantities in (\ref{itogeq1exin}) and (\ref{itogeq2exin}) can be
rewritten in standard physical units via dimensionless
parameters as
\begin{equation}
mr_h\to \mu\equiv\frac{mM}{M_{\rm Pl}^2}, \qquad qQ\to \nu\equiv\frac{qQ}{\hbar c}.
\end{equation}
We can also write $\mu=R_g/(2l_C)$, where $R_g$ is the
Schwarzschild radius and $l_C$ is the Compton wavelength of the particle \cite{GorNez12}. Similar to the previous case,
it is proven that for a regular solution
\begin{equation}
E=\frac{qQ}{r_h}.
\label{energyextr}
\end{equation}
Equations (\ref{itogeq1exin}) and (\ref{itogeq2exin}) when $y\to0$ then have
asymptotic solutions,
\begin{equation}
g=C_1y^\varkappa+C_2y^{-\varkappa}, \quad \mbox{where} \quad
\varkappa=\sqrt{k^2+\mu^2-\nu^2}.
\end{equation}
The solutions with $C_2=0$ are physical, because the
part of the solution at $C_2\neq0$ makes a diverging contribution to (12), similar to the case with non-extremal
black holes considered above. Then,
\begin{equation}
h=C_1y^\varkappa\frac{k+\varkappa}{\mu-\nu}.
\label{hc11}
\end{equation}
The contribution from (\ref{hc11}) to (\ref{normint}) is finite under the
condition $2\varkappa-2>-1$, which can be rewritten as
\begin{equation}
k^2+\mu^2-\nu^2>\frac{1}{4}.
\label{kmununon}
\end{equation}
Hence it follows that the state with $k=0$ is forbidden
if $1/4+\nu^2-\mu^2\geq0$. In view of the Pauli exclusion principle, each of the quantum levels found can be filled
with two identical fermions. A numerical solution of
Eqs.~(\ref{itogeq1}) and (\ref{itogeq2}) for the internal region $r<r_h$ is
shown in the figure~\ref{gr1extr}.

The irregular solution for $E\neq qQ/r_h$ and $y\to 0$ in
this case is
\begin{equation}
g=-C\sin(\alpha/y+\delta), \qquad h=C\cos(\alpha/y+\delta),
\label{nonreg2}
\end{equation}
where $\alpha=r_h(E-qQ/r_h)$. For this solution, $|g|^2+|h|^2=|C|^2$
and (\ref{normint}) diverges.

To investigate the case of $r\to r_h$ from the outside,
we will set $y=f^{1/2}=(1-r_h/r)$. The corresponding equations are 
self-consistent for the same value of (\ref{energyextr}) and
under condition (\ref{kmununon}) and have a physically acceptable
(with a finite integral (\ref{normint})) solution,
\begin{equation}
g=C_1y^\varkappa, \qquad h=C_1y^\varkappa\frac{k-\varkappa}{\mu+\nu}.
\label{hc12}
\end{equation}
Solutions (\ref{hc11}) and (\ref{hc12}) lose their meaning at $\mu=\pm\nu$,
respectively. Since the black hole is extremal, with
$M=|Q|$, the conditions $\mu=\pm\nu$ imply that the particle
itself is extremal, $m=|q|$. For ordinary particles, such
as the proton or electron, this equality, of course, does
not hold. For them, $m\ll|q|$, but the concept of extremality itself loses its meaning due to the contribution
of quantum effects. The solution of the Dirac equation
in the metric of a classical naked Reissner-Nordstr\"om
singularity with $|Q|>M$ was found numerically in
\cite{Dzh12}.

\section{Discussion} 

We showed that the stationary regular solution for a
fermion has the energy level $E=qQ/r_-$ below the
internal Cauchy horizon, where $q$ and $q$ are the particle and black hole charges, respectively, and $r_-$ is the
radius of the Cauchy horizon. The existence of only
one level is attributable to the properties of the wave
equation when the Cauchy horizon is approached
from the inside. There is also only one regular level
with energy $E=qQ/r_+$ outside the event horizon and
the solution at infinity corresponds to that for the
hydrogen atom. However, the probability integral
turned out to diverge on the horizons $r_\pm$ and, therefore,
the wave functions cannot be normalized to unity. This
may imply that the real solution tends to the stationary
solutions found only asymptotically, while the particle
tends to be localized on the horizons. If quantum tunneling is disregarded, then, as is well known, the falling
of a particle below the horizon, according to the clocks
of a remote observer, will take an infinitely long time.
Accordingly, the particle localization on the horizon
will occur only asymptotically on long time scales. In
the extremal case of $|Q|=M$, the energies of the internal and external levels take on identical values of$E=qQ/r_h$, where $r_h$ is the radius of the extremal black hole
horizon. In contrast to the case of a non-extremal
black hole, at $|Q|=M$ a physically acceptable solution
with normalized wave functions exists both inside
(below the Cauchy horizon) and outside the black
hole.

The stationary solutions of the Dirac equation we
considered show that black holes with charges on
inner quantum orbits can exist. An alternative possibility is a naked singularity around which charge is distribution in the stationary case \cite{Dzh12}. If $q=-Q$, where
$q$ is the total charge of the particles at quantum levels,
then the external metric of such a system is the
Schwarzschild one. For an external observer, the system is neutral and interacts weakly with the surrounding matter. If, in addition, such systems are stable
(which requires an additional study) and were born
effectively in the early Universe, then they could be
candidates for dark matter particles. Vronsky \cite{Vroetal13} considered a Dirac particle outside the horizon of a neutral Schwarzschild black hole and also came up with
the idea of such systems as dark matter particles. However, if the black hole is neutral and the particle is
charged, then the system carries a nonzero total electric charge and will interact via this charge with the
surrounding matter. In the case of a Reissner-Nordstr\"om black hole with $q=-Q$, the corresponding candidate dark matter particles are electrically neutral and
interact very weakly.

If the black hole has internal and external energy
levels, then quantum transitions between these levels
are possible. This allows the charge falling into a
Reissner-Nordstr\"om black hole to linger at a quantum level below the Cauchy horizon and probably to
tunnel outward or into the internal universe. The
quantum transition of a particle from the external stationary energy level (\ref{energyout}) to the corresponding internal
level (\ref{energy}) may be considered as an analog of the classical fall of a particle into a black hole. Since the energies at the external and internal levels for an extremal
black hole are equal, no energy is released during the
quantum transition. In the case of a non-extremal
black hole, energy can be released during the transitions between quasi-stationary levels:
\begin{eqnarray}
\Delta E&=&E_--E_+=qQ\left(\frac{1}{r_-}-\frac{1}{r_+}\right)=
\nonumber \\
&=&\frac{2Mc^2q}{Q}\sqrt{1-\frac{Q^2}{GM^2}},
\label{inouttrans}
\end{eqnarray}
where we restored the physical dimensions of the
quantities in the last equality. We see that much energy
can be radiated during such transitions for $q^2\sim Q^2\ll GM^2$. If the systems under consideration are dark matduce highly energetic particles that may contribute to
the ultrahigh energy cosmic rays. There will be no such
energy release for an extremal black hole, because $\Delta E=0$. These transitions may turn out to be fundamentally important at the final stage of Hawking black hole
evaporation, because they affect the phase volumes of
the final and intermediate states of the particles produced during evaporation. The existence of the quantum levels we detected is a necessary condition for
these quantum transitions, but the possibility of such
transitions itself, of course, requires an additional
study and justification. The transitions between levels,
including those through the horizon, can occur via
quantum tunneling with a probability proportional to
$\exp(-2{\rm Im} S)$, where the action $S=\int dr p_r(r)$ is calculated along a semiclassical trajectory \cite{Vol99}, \cite{MiaXueZha11}. Note
also the fundamental possibility of investigating the
internal structure of black holes using quantum tunneling between external and internal states.

This study was supported by the Ministry of Education and Science of the Russian Federation (contract 8525), grant NSh-871.2012.2, and 
RFBR~13-02-00257-a.


\begin{thebibliography}{99}

\bibitem{Fock29}V. Fock, {\it Zeitschrift fur Physik} {\bf 57}, 261 (1929); V.\,Fock, D.\,Iwanenko, {\it Zeitschrift fur Physik} {\bf 54}, 798 (1929).

\bibitem{Wei00} S. Weinberg, {\it Gravitation and Cosmology: Principles and
Applications of the General Theory of Relativity} (Wiley,
New York, 1972; Platon, Volgograd, 2000).

\bibitem{BriWhe57} D.R. Brill, J.A. Wheeler, {\it Reviews of Modern Physics} {\bf 29}, 465 (1957).

\bibitem{Muk00} B. Mukhopadhyay, {\it Class.~Quant.~Grav.} {\bf 17}, 2017 (2000).

\bibitem{Ber75} V.A. Berezin, Preprint: {\it Neutrino Forces and the
Schwarzschild Metric} (Institute for Nuclear Research,
Academy of Sciences of the Soviet Union, Moscow,
1975).

\bibitem{Bur08} A. Burinskii, {\it Grav. Cosmol.} {\bf 14}, 109 (2008).

\bibitem{DerRuf74} N. Deruelle, R. Ruffini, {\it Physics Letters} B {\bf 52},  437 (1974).
 
\bibitem{Teretal78} I.M.~Ternov, V.R.~Khalilov, G.A.~Chizhov, and
A.B.~Gaina, {\it Sov. Phys. J.} {\bf 21} (9), 1200 (1978).

\bibitem{Kof82} L.A. Kofman, {\it Phys.~Lett.} A {\bf 87}, 281 (1982).

\bibitem{SofMulGre77} M. Soffel, B. Muller, and W. Greiner, {\it J. Phys. A: Math. Gen.} {\bf 10}, 551 (1977).

\bibitem{Teretal80} I.M.~Ternov, A.B.~Gaina, and G.A.~Chizhov, {\it Sov.
Phys. J.} {\bf 23} (8), 695 (1980).

\bibitem{GalPomChi83} D.V. Galtsov, G.V. Pomerantseva, and G.A. Chizhov,
{\it Sov. Phys. J.} {\bf 26} (8), 743 (1983).

\bibitem{GorNez12} M.V. Gorbatenko, V.P. Neznamov, arXiv:1205.4348 [gr-qc].

\bibitem{Vroetal13} M.A. Vronsky et al., arXiv:1301.7595 [gr-qc].

\bibitem{Dzh12} V. Dzhunushaliev, arXiv:1202.5100 [gr-qc].

\bibitem{chandra} S. Chandrasekhar, {\it The Mathematical Theory of Black
Holes} (Oxford University Press, Oxford, 1983; Mir,
Moscow, 1986), Part 1, Chap. 5.

\bibitem{Dok11} V.I. Dokuchaev, {\it Class. Quant. Grav.} {\bf 28}, 235015 (2011).

\bibitem{BalBicStul89} J.\,Bi\v{c}\'ak, Z.\,Stuchl{\'\i}k  and V.\,Balek  
{\it Bull. Astron. Inst. Czechosl.} {\bf 40}, 65 (1989); {\sl ibid.} {\bf 40}, 133 (1989).

\bibitem{Kagramanova10} E. Hackmann et al., {\it Phys. Rev.} D {\bf 81} 044020 (2010).

\bibitem{GrunKag10} S.\,Grunau and V.\,Kagramanova, {\it Phys. Rev.} D {\bf 83}, 044009 (2011)

\bibitem{Olietal11} M.\,Olivares et al., arXiv:1101.0748 [gr-qc].

\bibitem{CarGilLid94} B.J. Carr, J.H. Gilbert and J.E. Lidsey,
{\it Phys. Rev.} D {\bf 50}, 4853 (1994)

\bibitem{DolNasNov00} A.D. Dolgov, P.D. Naselsky and I.D. Novikov, arXiv:astro-ph/0009407.

\bibitem{Carr03} B.J. Carr,	{\it Lect. Notes Phys.} {\bf 631}, 301 (2003); arXiv:astro-ph/0310838.

\bibitem{Mar66} M.A. Markov, {\it Sov. Phys. JETP} {\bf 24} (3), 584 (1966).

\bibitem{Vol99} G.E. Volovik, {\it JETP Lett.} {\bf 69} (9), 705 (1999).

\bibitem{MiaXueZha11} Y. Miao, Z. Xue, and S. Zhang, {\it Europhys.~Lett.} {\bf 96}, 10008 (2011).

\end{thebibliography}
\end{document}